\documentclass[aps,pra,twocolumn,showpacs,groupedaddress, nofootinbib]{revtex4-1}
\usepackage{graphicx}
\usepackage{float}
\usepackage{longtable}

\begin{document}

%\preprint{}

\title{Production of strongly bound $^{39}$K bright solitons}


\author{S. Lepoutre, L. Fouch\'e, A. Boiss\'e, G. Berthet, G. Salomon, A. Aspect, T. Bourdel}
%\thanks{}
\affiliation{Laboratoire Charles Fabry, Institut d'Optique, CNRS, Univ. Paris-Sud, 2, Avenue Augustin Fresnel, 91127 PALAISEAU Cedex, France}


\date{\today}

\begin{abstract}
We report on the production of $^{39}$K matter-wave bright solitons, {\it i.e.}, 1D matter-waves that propagate without dispersion thanks to attractive interactions. The volume of the soliton is studied as a function of the scattering length through three-body losses, revealing peak densities as high as 
$\sim 5 \times10^{20}$\,m$^{-3}$. Our solitons, close to the collapse threshold, are strongly bound and will find applications in fundamental physics and atom interferometry.

\end{abstract}

\pacs{03.75.Lm,  67.85.-d}

%\keywords{}

\maketitle


Solitons are one-dimensional wave-packets that propagate with neither change of shape nor loss of energy. They are a consequence of non-linearities that balance wave-packet spreading due to dispersion.  They appear in numerous physical systems such as water waves, optical fibers, plasmas, acoustic waves or even in energy propagation along proteins \cite{Malomed2005}. Solitons are also observed in ultracold quantum gases \cite{Burger1999, Denschlag2000, Khaykovich2002, Strecker2002, Eiermann2004}. In this context, matter-wave bright solitons are Bose-Einstein condensates that  remain bound thanks to mean-field attractive interactions in a one dimensional geometry \cite{Khaykovich2002, Strecker2002}.

Matter-wave bright solitons are predicted to be a great tool to locally probe rapidly varying forces for example close to a surface \cite{Ernst2010, Cornish2009}, or probe (surface) bound states \cite{Ernst2010, Damon2016} which do not appear in linear scattering. For example, the small size of bright solitons has been used in the measurement of quantum reflection from a barrier \cite{Marchant2013, Marchant2016}. Because of their dispersion-free propagation, bright solitons are also believed to be good candidates for performing very long time atom interferometry measurements \cite{Cronin2009} although interactions may cause additional phase shifts \cite{Billam2012, Martin2012,  Polo2013, Helm2015}. Recently, an experiment demonstrated an increased visibility for a soliton atomic interferometer as compared to its non interacting counterpart \cite{McDonald2014}. The interactions in solitons can also lead to squeezed or entangled states, which could improve the sensitivity of interferometric measurements beyond the shot noise limit \cite{Jo2007, Veretenov2007, Lee2012, Kasevich2012, Gertjerenken2013, Helm2014, Gertjerenken2015}.  In some cases, the formation of mesoscopic Schr\"odinger cat states or NOON states is predicted  \cite{Weiss2009, Streltsov2009, Streltsov2009b}. A problem in using these states is losses, such as three-body collisions, which are an intrinsic source of decoherence. They can also induce unusual soliton center of mass dynamics \cite{Weiss2015}. 

\begin{figure}
\centering
\includegraphics[width=0.5\textwidth]{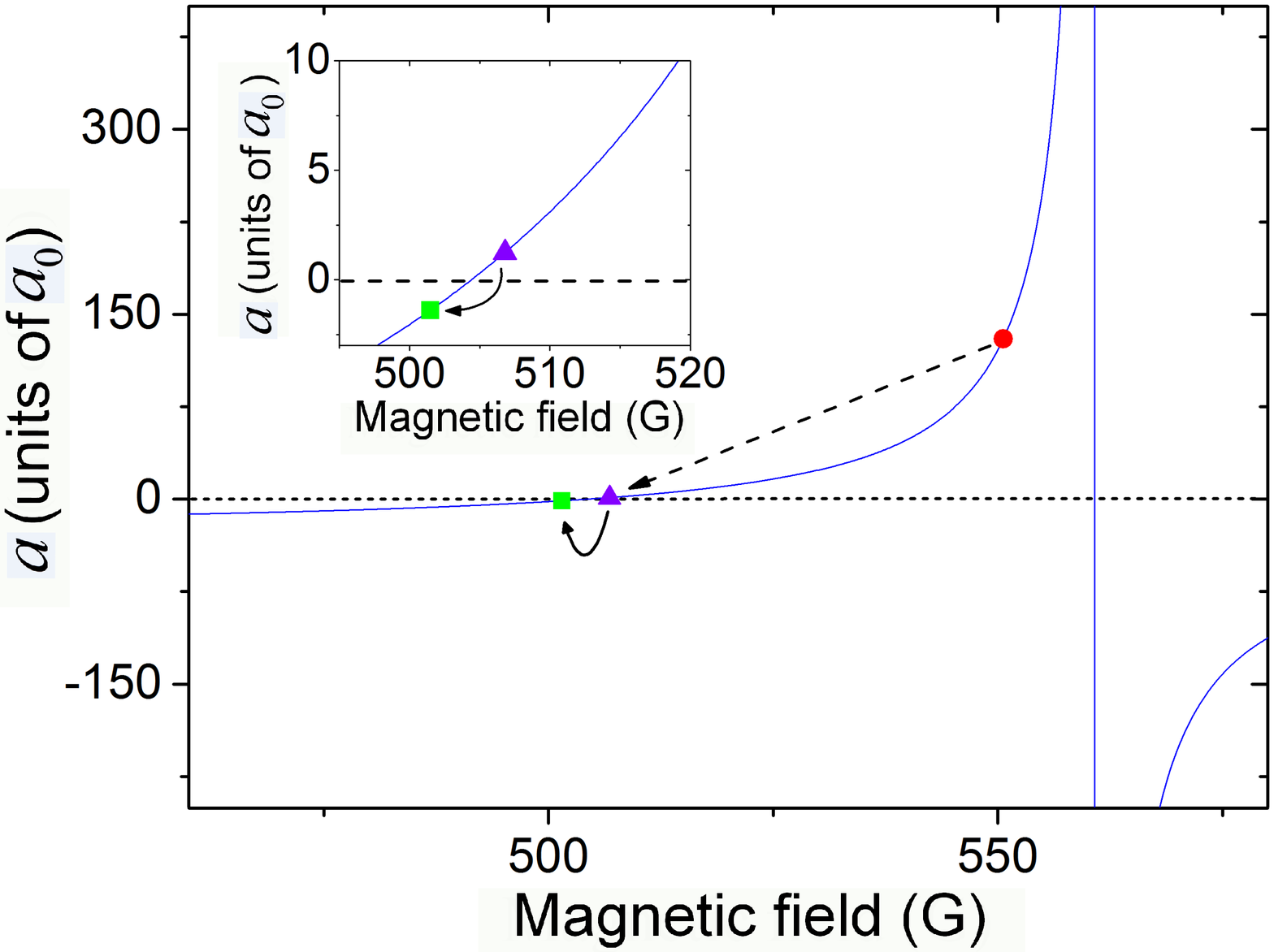} 
\caption{(Color online) Scattering length as a function of the magnetic field for $^{39}$K in the $|F=1, m_F=-1\rangle$ state \cite{Derrico2007}. Inset: Zoom around the zero-crossing of the scattering length. The evaporation to Bose-Einstein condensation takes place at 550\,G (red bullet). The magnetic field is then ramped in two steps to 507\,G (violet triangle) and then to 501.3\,G (green square) where the scattering length is -1.5\,$a_0$ in order to produce bright solitons.}
\label{Feshbach}
\end{figure}

Experiments producing and studying matter-wave bright solitons, despite their interest in both applied and fundamental physics, have remained scarce. In fact, only two elements have been turned into bright solitons, $^7$Li \cite{Khaykovich2002, Strecker2002, Medley2014, Nguyen2014} and $^{85}$Rb \cite{Cornish2006, Marchant2013}.  In this paper, we describe the production of $^{39}$K solitons in the $|F=1, m_F=-1\rangle$ state using the Feshbach resonance at 561\,G \cite{Derrico2007} and its associated zero-crossing of the scattering length at 504.4\,G (see figure \ref{Feshbach}).  We have optimized the setup in order to produce strongly bound solitons, i.e., solitons with a large negative interaction energy. We thus produce very dense solitons close to the threshold for collapse \cite{Khaykovich2002, Strecker2002}. We observe significant three-body losses with peak density up to $5\times 10^{20}$\,m$^{-3}$. We study the three-body loss rate as a function of the scattering length $a$ on both sides of the zero-crossing. The observed strong variations of  loss rates are well explained by a simple mean-field model that predicts variations of the size of the condensates or solitons and assumes a constant three-body loss coefficient $K_3$, yielding $K_3=1.5(6)\times 10^{-41}$m$^6$.s$^{-1}$. We are able to reach a regime where the interaction energy of the soliton exceeds its center of mass kinetic energy, and where the atoms are predicted to behave collectively in scattering \cite{Lee2006, Hansen2012,  Gertjerenken2012, Helm2012, Cuevas2013, Helm2014}.

%\begin{figure}
%\centering
%\includegraphics[width=0.35\textwidth]{FORTs.png} 
%\caption{Top view of the different optical trapping beams used in the experiment. The beams are all horizontal. The angles and the relative sizes are respected} 
%\label{Feshbach}
%\end{figure}

The creation of potassium bright solitons is based on the all-optical production of $^{39}$K Bose-Einstein condensates \cite{Salomon2014b}. A crucial ingredient allowing an efficient direct loading of the optical trap is the gray molasses cooling of potassium \cite{Salomon2014a, Nath2013}. Evaporative cooling is performed at 550\,G, in the wing of the Feshbach resonance, where the scattering length is 130\,$a_0$ with $a_0$ the Bohr radius (see figure \ref{Feshbach}). The final trap is a far off resonance optical dipole trap made from two horizontal crossing beams. The first one at 1064\,nm with a waist of 48.5\,$\mu$m (radius at $1/e^{2}$) permits a strong radial confinement while the second one at 1550\,nm with a waist of 150\,$\mu$m is used to provide a weak longitudinal confinement (44\,Hz). The final evaporation is performed by lowering the power of the 1064\,nm beam down to 56\,mW such that the most energetic atoms fall under gravity. We obtain almost pure condensates with up to 4$\times 10^4$ atoms. The radial trap is then recompressed up to a  power of 117\,mW to form an elongated trap, whose frequencies are measured through parametric oscillations to be 195\,Hz$\times$195\,Hz$\times$44\,Hz.

\begin{figure}
\centering
\includegraphics[width=0.45\textwidth]{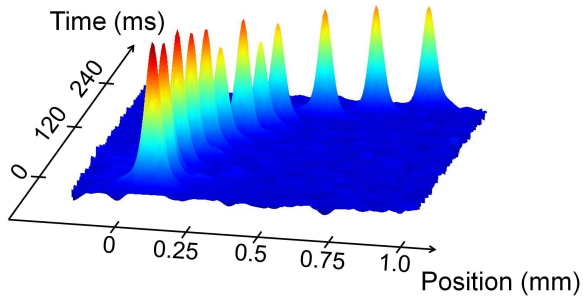} 
\caption{(Color online) Density profiles of solitons as a function of time. Images are separated by 20\,ms and stack vertically. The acceleration at the release point is $5$\,mm.s$^{-2}$.} 
\label{solitons}
\end{figure}

The final step to produce solitons consists of modifying the scattering length by changing the magnetic field value. This is done in two steps, first to 507\,G in 150\,ms approaching the zero-crossing from the positive side and then to 501.3\,G in 400\,ms, where the scattering length $a=-1.5(2)\,a_0$ is then negative (see figure \ref{Feshbach}). The condensate then shrinks and forms the solitons. The ramp times are relatively long compared to the inverse of the longitudinal trapping frequency, preventing the condensate from being excited. Figure \ref{solitons} shows the propagation of solitons in the 1064\,nm optical trap, when the longitudinal confining beam is switched off. The longitudinal potential has been characterized in detail. It has an anti-trapping curvature ($i \times 1.9\,$Hz) which mainly originates from the bias magnetic field curvature. The acceleration at the release point can be varied at will by introducing a weak magnetic field gradient along the trapping beam with an auxiliary coil. In figure \ref{solitons}, we observe the characteristic absence of dispersion for the solitons during the 250\,ms propagation time. Their center of mass is moving by about 1\,mm along an hyperbolic trajectory because of a $5$\,mm.s$^{-2}$ acceleration at the release point. 

Images are taken by fluorescence imaging after the following sequence. The optical trap is first switched off abruptly. After 7\,ms of expansion, the magnetic field is switched off. At this time, the gas is already in a ballistic regime and is sufficiently diluted to avoid losses while crossing the lower field Feshbach resonances.  An additional delay of 15\,ms permits the eddy currents to damp. The four horizontal beams from the magneto-optical trap cooling laser, tuned to be on resonance with the optical transitions, are then shined on the atoms and their fluorescence signal is collected from above during 100\,$\mu$s and recorded with an EMCCD camera (Andor iXon). The duration of the imaging pulse is chosen to optimize the signal without introducing too high blurring. The overall resolution is then 15\,$\mu$m, which exceeds the in-situ micrometer size of solitons as well as their size after 22\,ms expansion. Over the 250\,ms of propagation, the longitudinal sizes of the solitons are given by this resolution limit.

The initial atom number in our solitons is typically 6$\times 10^3$, a number which is well below the initial condensate atom number \cite{number}. Actually, the atom number also decreases by an additional 25$\%$ during the 250\,ms propagation time. This is a consequence of three-body losses whose rate  increases with the density when the scattering length is reduced toward zero or negative values, and which will be studied in more details below. Note that such important three-body losses lead to a stabilization of the atom number in the solitons and we see no significant difference in soliton atom number when the initial atom number is decreased by a factor two. 

The calibration of the scattering lengths is based on the measurement of the longitudinal expansion of a condensate when varying the current flowing through the Feshbach field coils. In practice, the zero-crossing of the scattering length is spotted when the longitudinal expansion of the gas corresponds to the one of a condensate, interacting solely via the dipole-dipole interaction (whose effect is small although non-negligible in our case) \cite{Lahaye2009}. We then rely on the scattering model from \cite{Derrico2007}, to deduce all magnetic field values and their corresponding scattering length. The scattering lengths are calibrated with an accuracy of 0.2\,$a_0$ in the region of interest, {\it i.e.}, close to the zero-crossing. 

We observe the non dispersive propagation of solitons only in a relatively narrow region of scattering lengths. For $a\geq -0.9(2)\,a_0$, the condensate expands because of the initial confinement energy. For $a\leq -2.15(20)\,a_0$, we observe a collapse. With about 4.5$\times 10^3$ atoms this corresponds to a value of the parameter $N |a|/\sigma_\rho$=0.45(10), where $\sigma_\rho=\sqrt{\hbar/m \omega_\perp}$ is the radial harmonic oscillator length. Theoretically, the limit of stability is  $N |a|/\sigma_\rho$=0.627 in the absence of longitudinal confining potential \cite{Carr2002} . Our observed slightly smaller value can be explained by important three-body losses during the formation of the soliton and prior to its observation close to the collapse (see below). Note that, when we encounter a collapse, we observe the disappearance of all condensed atoms. This is in contrast with recent experiments done in 3D Bose-gases trapped in a box potential \cite{Hadzibabic2016}.

\begin{figure}
\centering
\includegraphics[width=0.5\textwidth]{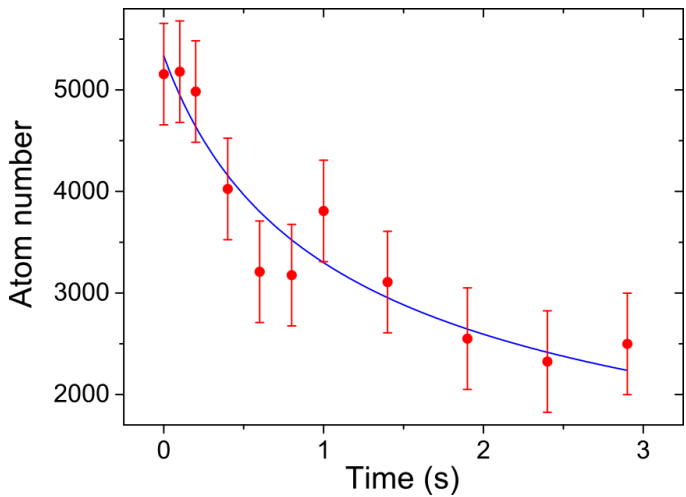} 
\caption{(Color online) Three-body loss curve as a function of time at 502.8\,G corresponding to a=-0.8\,$a_0$. The solid line corresponds to the fit using equation \ref{eq:decay}. The error bars are r.m.s. shot to shot variations. }
\label{decay}
\end{figure}
\begin{figure}
\centering
\includegraphics[width=0.5\textwidth]{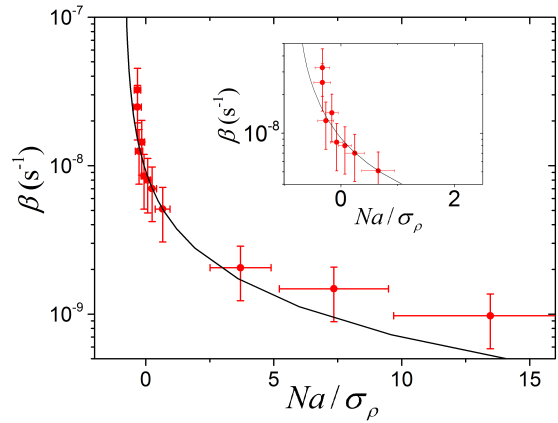} 
\caption{(Color online) Three-body loss rate coefficient $\beta$ as a function of the dimensionless parameter $N a/\sigma_\rho$. The solid line corresponds to the variational theory described in the text. Inset: Zoom on the zero-crossing region. The error bars along both axis include the systematic uncertainty on the calibration of the atom number.}
\label{beta}
\end{figure}

We now study the losses as a function of the magnetic field or equivalently as a function of the scattering length. We focus our study in the region where the scattering length is varied from 24\,$a_0$ to $-2\,a_0$. For a fixed value of the magnetic field, we observe the decrease of the atom number as a function of a waiting time at the end of the preparation sequence. A typical decay curve is plotted in figure \ref{decay}. We intentionally stop our analysis after 3 seconds such that most atoms remain in the condensate and not to be fooled by thermal atoms. On this time scale, it is difficult to discriminate the nature of losses. We have measured a 30\,s one-body decay time, and such losses are negligible. As we are not in the absolute ground state, two-body relaxations are energetically allowed. Nevertheless, they do not conserve the total spin and require dipole-dipole interaction. Their rate is thus expected to be small as compared to the three-body loss rate \cite{Shotan2014}. The atom decay curves are thus experimentally fitted using the loss equation 
\begin{equation}
\dot{N}=-\beta N^3 \label{eq:decay}
\end{equation}
with constant $\beta$. We observe in figure \ref{beta} that the fitted $\beta$ coefficient strongly varies as a function of the dimensionless parameter $N a/\sigma_\rho$.  As we explore only a small region in magnetic field, the variation of the loss rate is not likely to be a consequence of a variation of the loss rate coefficient $K_3$ but rather a consequence of the variation of the effective volume of the condensate, and thus of the density, when changing the interaction parameters \cite{Shotan2014}. An increase by a factor of 30 in $\beta$, as shown in figure 4, corresponds to an increase by a factor of 5.5 in the density, for a constant loss rate coefficient. The effective volume of the condensate continues to decrease when the scattering is tuned from zero to negative values, as expected for a soliton. 

Our experimental measurements of the three-body coefficients can be compared to the expectations from the mean-field theory. In practice, we use a cylindrical gaussian ansatz wave function, which is known to give a good estimate of the density profile \cite{Billam2012} all the way from positive to negative scattering lengths.
\begin{equation}
\psi(\textrm{\bf r})=\frac{1}{(2\pi)^{1/2} \sigma_r} \exp(-{r}^2/4 \sigma_r^2) \frac{1}{(2 \pi)^{1/4}\sigma_z^{1/2}}\exp(-z^2/4\sigma_z^2)
\end{equation}
The r.m.s. sizes  $\sigma_r$ and $\sigma_z$ are variational parameters that we use to minimize the Gross-Pitaevskii energy functional as a function of the interaction parameter $Na$ \cite{Khaykovich2002, Carr2002, Billam2012}. An integration over the density profile gives $\beta=K_3/(2\pi)^3/3^{3/2}/\sigma_r^4/\sigma_z^2$, where $K_3$ is the condensate three-body loss coefficient. In figure \ref{beta}, this simple approximate theory is found to reproduce fairly well the data with a constant value of $K_3$ furthermore validating our assumption of a dominant three-body loss mechanism. For the highest values of $a$, we observe a deviation from the theoretical curve that we attribute to an increase of the three-body coefficient as we move toward the Feshbach resonance. We find $K_3$=$1.5(6)\times 10^{-41}$ m$^{6}$.s$^{-1}$ in the region of the zero-crossing where the uncertainty mostly comes from the atom number calibration. This is comparable to the value of $1.3(5)\times 10^{-41}$ m$^{6}$.s$^{-1}$ measured in the absolute ground state of potassium 39 around its zero-crossing \cite{Fattori2008}. Both values have the right order of magnitude for a non-resonant three-body loss coefficient expected from the Van der Waals coefficient of potassium \cite{Shotan2014}. Comparisons with $^7$Li and $^{85}$Rb solitons are difficult as the loss rates close to the zero-crossings are not well documented. 

In the case of the densest solitons, obtained at the lowest values of $a$, we can infer a high peak density of $\sim 5\times10^{20}$ \,m$^{-3}$. This corresponds to an interaction energy per particle of 30\,Hz in the Gross-Pitaevskii energy functional  \cite{Khaykovich2002, Carr2002, Billam2012}. This value can be compared to the center of mass kinetic energy in our soliton. Experimentally, we measure a shot to shot fluctuation of the position of the solitons after a  propagation time of 200\,ms, corresponding to an r.m.s. initial velocity fluctuation of 0.15\,mm/s. This fluctuation probably originates from residual dipole oscillation in the trap. Such a velocity corresponds to a kinetic energy per atom of about 1\,Hz. We can thus produce solitons in the interesting situation where the interaction energy of the soliton dominates over its kinetic energy. In this regime, the atoms are expected to behave collectively, for example in the collision with a potential barrier \cite{Lee2006, Hansen2012,  Gertjerenken2012, Helm2012, Cuevas2013, Helm2014}.

More generally, our work opens a new experimental platform to study the matter-wave bright solitons both for fundamental and applied physics. Atom interferometry with solitons is certainly worth investigating. Further studies include the soliton dynamics after a quench of one of the parameters such as the scattering length \cite{Franchini2015, Gamayun2015}. Relaxation in such an interacting quantum integrable system with attractive interaction is of particular interest \cite{Polkovnikov2011, Cazalilla2011, Goldstein2014}. Another interesting direction would be to experimentally produce liquid droplets that are predicted to form in a Bose-Bose mixture because of a compensation between two-body mean-field interaction and repulsive three-body interaction \cite{Petrov2014} or beyond mean-field corrections \cite{Petrov2015, Petrov2016}. A mixture of $^{39}$K in two different spin states has been predicted to be especially suited for these studies \cite{Petrov2014, Petrov2015}. Similar droplets have recently been observed in dipolar condensates \cite{Kadau2016, Ferrier2016, Chomaz2016}.

\begin{acknowledgments}
We thank K. Lumer for his experimental contribution. 
This research has been supported by CNRS, Minist\`ere de l'Enseignement Sup\'erieur et de la Recherche, Direction G\'en\'erale de l'Armement, ANR-12-BS04-0022-01, Labex PALM, ERC senior grant Quantatop, R\'egion Ile-de-France in the framework of DIM Nano-K, EU - H2020 research and innovation program (Grant No. 641122 - QUIC), Triangle de la physique. LCF is member of IFRAF.
\end{acknowledgments}

%\bibliography{biblio}

\begin{thebibliography}{0}%
\makeatletter
\providecommand \@ifxundefined [1]{%
 \@ifx{#1\undefined}
}%
\providecommand \@ifnum [1]{%
 \ifnum #1\expandafter \@firstoftwo
 \else \expandafter \@secondoftwo
 \fi
}%
\providecommand \@ifx [1]{%
 \ifx #1\expandafter \@firstoftwo
 \else \expandafter \@secondoftwo
 \fi
}%
\providecommand \natexlab [1]{#1}%
\providecommand \enquote  [1]{``#1''}%
\providecommand \bibnamefont  [1]{#1}%
\providecommand \bibfnamefont [1]{#1}%
\providecommand \citenamefont [1]{#1}%
\providecommand \href@noop [0]{\@secondoftwo}%
\providecommand \href [0]{\begingroup \@sanitize@url \@href}%
\providecommand \@href[1]{\@@startlink{#1}\@@href}%
\providecommand \@@href[1]{\endgroup#1\@@endlink}%
\providecommand \@sanitize@url [0]{\catcode `\\12\catcode `\$12\catcode
  `\&12\catcode `\#12\catcode `\^12\catcode `\_12\catcode `\%12\relax}%
\providecommand \@@startlink[1]{}%
\providecommand \@@endlink[0]{}%
\providecommand \url  [0]{\begingroup\@sanitize@url \@url }%
\providecommand \@url [1]{\endgroup\@href {#1}{\urlprefix }}%
\providecommand \urlprefix  [0]{URL }%
\providecommand \Eprint [0]{\href }%
\providecommand \doibase [0]{http://dx.doi.org/}%
\providecommand \selectlanguage [0]{\@gobble}%
\providecommand \bibinfo  [0]{\@secondoftwo}%
\providecommand \bibfield  [0]{\@secondoftwo}%
\providecommand \translation [1]{[#1]}%
\providecommand \BibitemOpen [0]{}%
\providecommand \bibitemStop [0]{}%
\providecommand \bibitemNoStop [0]{.\EOS\space}%
\providecommand \EOS [0]{\spacefactor3000\relax}%
\providecommand \BibitemShut  [1]{\csname bibitem#1\endcsname}%
\let\auto@bib@innerbib\@empty
%</preamble>
\end{thebibliography}%


\begin{thebibliography}{00}

\bibliographystyle{apsrev4-1}

\bibitem{Malomed2005}
B. Malomed, in {\it Encyclopedia of Nonlinear Science}, edited by A. Scott (Routledge, New York, 2005) pp. 639-643.

\bibitem{Khaykovich2002}
L. Khaykovich, F. Schreck, G. Ferrari, T. Bourdel, J. Cubizolles, L.D. Carr, Y. Castin, C. Salomon, Science {\bf 296}, 1290 (2002).

\bibitem{Strecker2002}
K.E. Strecker, G.B. Partridge, A.G. Truscott, R.G. Hulet, R.G., Nature {\bf 417}, 150 (2002).

\bibitem{Burger1999}
S. Burger, K. Bongs, S. Dettmer, W. Ertmer, K. Sengstock, A. Sanpera, G.V. Shlyapnikov, and M. Lewenstein, Phys. Rev. Lett. {\bf 83}, 5198 (1999).

\bibitem{Denschlag2000}
J. Denschlag, J. E. Simsarian, D. L. Feder, C.W. Clark,
L. A. Collins, J. Cubizolles, L. Deng, E.W. Hagley,
K. Helmerson, W.P. Reinhardt, S.L. Rolston, B.I. Schneider,
W.D. Phillips, Science {\bf 287}, 97 (2000).


\bibitem{Eiermann2004}
B. Eiermann, Th. Anker, M. Albiez, M. Taglieber, P. Treutlein, K.-P. Marzlin, and M.K. Oberthaler,
Phys. Rev. Lett. {\bf 92}, 230401 (2004).


\bibitem{Ernst2010}
T. Ernst and J. Brand,
Phys. Rev. A {\bf  81}, 033614 (2010).


\bibitem{Cornish2009}
S.L. Cornish, N.G. Parker, A.M. Martin, T.E. Judd, R.G. Scott, T.M. Fromhold, C.S. Adams,
Physica D: Nonlinear Phenomena {\bf 238}, 1299  (2009).

\bibitem{Damon2016}
F. Damon, B. Georgeot, D. Gu\'ery-Odelin, Europhysics Lett.{\bf 115}, 20010 (2016).

\bibitem{Marchant2013}
A.L. Marchant, T.P. Billam, T.P. Wiles, M.M.H. Yu, S.A. Gardiner, S.L. Cornish,
Nature Communications {\bf 4}, 1865 (2013).

\bibitem{Marchant2016}

A.L. Marchant, T.P. Billam, M.M.H. Yu, A. Rakonjac, J.L. Helm, J. Polo, C. Weiss, S.A. Gardiner, and S.L. Cornish,
Phys. Rev. A {\bf 93}, 021604(R) (2016).

\bibitem{Cronin2009}
A.D. Cronin, J. Schmiedmayer, and D.E. Pritchard,
Rev. Mod. Phys. {\bf 81}, 1051 (2009).



\bibitem{Helm2015}
J. L. Helm, S. L. Cornish, and S. A. Gardiner
Phys. Rev. Lett. {\bf 114}, 134101 (2015). 


\bibitem{Martin2012}
A.D. Martin, J. Ruostekoski, New Journal of physics {\bf 14}, 043040  (2012).

\bibitem{Polo2013}
J. Polo and V. Ahufinger,
Phys. Rev. A {\bf 88}, 053628 (2013).


\bibitem{Billam2012}
T.P. Billam, A.L. Marchant, S.L. Cornish, S.A. Gardiner, N.G. Parker, Chapter in "Spontaneous Symmetry Breaking, Self-Trapping, and Josephson Oscillations", edited by Boris Malomed (Springer, 2013)
arXiv:1209.0560 (2012).


\bibitem{McDonald2014}
G.D. McDonald, C.C.N. Kuhn, K.S. Hardman, S. Bennetts, P.J. Everitt, P.A. Altin, J.E. Debs, J.D. Close, and N.P. Robins
Phys. Rev. Lett. {\bf 113}, 013002 (2014).


\bibitem{Helm2014}
J.L. Helm, S.J. Rooney, C. Weiss, and S.A. Gardiner,
Phys. Rev. A {\bf 89}, 033610 (2014).

\bibitem{Gertjerenken2013}
B. Gertjerenken,
Phys. Rev. A {\bf 88}, 053623 (2013).

%\bibitem{Fiorentino2001}
%Soliton squeezing in a Mach-Zehnder fiber interferometer
%Marco Fiorentino, Jay E. Sharping, Prem Kumar, Dmitry Levandovsky, and Michael Vasilyev
%Phys. Rev. A 64, 031801(R) ? Published 3 August 2001


\bibitem{Kasevich2012}
M. Kasevich, Atom systems and Bose-Einstein condensates for metrology and navigation, first NASA Quantum Future Technologies conference (2012):http://quantum.nasa.gov/materials/2012-01-18-B1-Kasevich.pdf

\bibitem{Lee2012}
C. Lee, J.H. Huang, H.M. Deng, H.Dai, J. Xu, Frontiers of physics {\bf 7}, 109 (2012).


%Absolute Geodetic Rotation Measurement Using Atom Interferometry
%J. K. Stockton, K. Takase, and M. A. Kasevich
%Phys. Rev. Lett. 107, 133001 ? Published 22 September 2011

%M. Fattori, C. D'Errico, G. Roati, M. Zaccanti, M. Jona-
%Lasinio, M. Modugno, M. Inguscio, and G. Modugno,
%Phys. Rev. Lett. 100, 080405 (2008)

%C. Chin, R. Grimm, P. Julienne, and E. Tiesinga, Rev.
%Mod. Phys. 82, 1225 (2010).

%Matter-wave soliton interferometer based on a nonlinear splitter
%By:Sakaguchi, H (Sakaguchi, Hidetsugu)[ 1 ] ; Malomed, BA (Malomed, Boris A.)[ 2 ]
%NEW JOURNAL OF PHYSICS
%Volume: 18
%Article Number: 025020
%DOI: 10.1088/1367-2630/18/2/025020
%Published: FEB 22 2016

\bibitem{Veretenov2007}
N. Veretenov, Y. Rozhdestvensky, N. Rosanov, V. Smirnov, S. Fedorov,
The European Physical Journal D {\bf 42}, 455 (2007).

\bibitem{Jo2007}
G.-B. Jo, Y. Shin, S. Will, T.A. Pasquini, M. Saba, W. Ketterle, D.E. Pritchard, M. Vengalattore, and M. Prentiss
Phys. Rev. Lett. {\bf 98}, 030407 (2007).

\bibitem{Gertjerenken2015}
B. Gertjerenken, T.P. Wiles, C. Weiss,
arXiv:1508.00656


\bibitem{Weiss2009}
C. Weiss and Y. Castin,
Phys. Rev. Lett. {\bf 102}, 010403 (2009).

\bibitem{Streltsov2009}
A.I. Streltsov, O.E. Alon, and L.S. Cederbaum
Phys. Rev. A {\bf 80}, 043616 (2009).

\bibitem{Streltsov2009b}
A.I. Streltsov, O.E. Alon, and L.S. Cederbaum, J. Phys. B {\bf 42}, 091004 (2009).


\bibitem{Weiss2015}
C. Weiss, S.A. Gardiner, and H.-P. Breuer,
Phys. Rev. A {\bf 91}, 063616 (2015).


\bibitem{Medley2014}
P. Medley, M.A. Minar, N.C. Cizek, D. Berryrieser, and M.A. Kasevich
Phys. Rev. Lett. {\bf 112}, 060401 (2014).


\bibitem{Nguyen2014}
J.H.V. Nguyen, P. Dyke, D. Luo, B. Malomed, and R.G.
Hulet, Nat. Phys. {\bf 10}, 918 (2014).


\bibitem{Cornish2006}
S.L. Cornish, S.T. Thompson, and C. E. Wieman, Phys. Rev.
Lett. {\bf 96}, 170401 (2006).

\bibitem{Derrico2007}
C. D'Errico, M. Zaccanti, M. Fattori, G. Roati, M. Inguscio, G. Modugno and A. Simoni, New Journal of physics {\bf 9}, 223 (2007).

\bibitem{Gertjerenken2012}
B. Gertjerenken, T.P. Billam, L. Khaykovich, and C. Weiss
Phys. Rev. A {\bf 86}, 033608 (2012).




%Nonlinear transport of Bose-Einstein condensates through waveguides with disorder
%By: Paul, T; Leboeuf, P; Pavloff, N; et al.
%PHYSICAL REVIEW A  Volume: 72   Issue: 6     Article Number: 063621   Published: DEC 2005




\bibitem{Lee2006}
C. Lee and J. Brand,
Europhysics Lett. {\bf 73}, 321 (2006).

\bibitem{Hansen2012}
S.D. Hansen, N. Nygaard and K. M\o{}lmer, arXiv:1210.1681

%Friction and diffusion of matter-wave bright solitons
%By:Sinha, S (Sinha, S); Cherny, AY (Cherny, AY); Kovrizhin, D (Kovrizhin, D); Brand, J (Brand, J)
%View ResearcherID and ORCID
%PHYSICAL REVIEW LETTERS
%Volume: 96  Issue: 3
%Article Number: 030406
%DOI: 10.1103/PhysRevLett.96.030406
%Published: JAN 27 2006



%Variational determination of approximate bright matter-wave soliton solutions in anisotropic traps
%T. P. Billam, S. A. Wrathmall, and S. A. Gardiner
%Phys. Rev. A 85, 013627 ? Published 19 January 2012

\bibitem{Helm2012}
J.L. Helm, T.P. Billam, and S.A. Gardiner,
Phys. Rev. A {\bf 85}, 053621 (2012).

\bibitem{Cuevas2013}
J.Cuevas-Maraver, P.G. Kevrekidis, B.A. Malomed, P. Dyke, and R.G. Hulet, New Journal of Physics {\bf 15}, 063006 (2013).

%\bibitem{Ruprecht1995}
%P. A. Ruprecht, M. J. Holland, K. Burnett, and Mark Edwards
%Phys. Rev. A {\bf 51}, 4704 (1995).

%\bibitem{Gerton2000}
%J.M. Gerton, D. Strekalov, I. Prodan, and R.G. Hulet, Nature {\bf 408}, 692 (2000).

%\bibitem{Billam2012b}
%T.P. Billam, S.A. Wrathmall, and S.A. Gardiner
%Phys. Rev. A {\bf 85}, 013627 (2012).





%Kasevich, M. A. Atom interferometry with Bose-Einstein condensed
%atoms. Comptes Rendus de l Academie des Sciences Serie IV,
%Physique et Astrophysique 2, 497-507 (2001).



\bibitem{Salomon2014b}
G. Salomon, L. Fouch\'e, S. Lepoutre, A. Aspect, and T. Bourdel, Phys. Rev. A {\bf 90}, 033405 (2014).

\bibitem{Salomon2014a}
G. Salomon, L. Fouch\'e, P. Wang, A. Aspect, P. Bouyer, T. Bourdel, Europhys. Lett. {\bf 104}, 63002 (2014).

\bibitem{Nath2013}
D. Nath, R.K. Easwaran, G. Rajalakshmi, and C.S.
Unnikrishnan, Physical Review A {\bf 88}, 053407 (2013).

\bibitem{number}
{The atom number is calibrated through the observation of the condensation threshold in a non interacting gas, with a 15\,$\%$ accuracy.}

\bibitem{Lahaye2009}
T. Lahaye, C. Menotti, L. Santos, M. Lewenstein and T. Pfau,
Reports on Progress in Physics {\bf 72}, 126401 (2009).

\bibitem{Carr2002}
L.D. Carr and Y. Castin,
Phys. Rev. A {\bf 66}, 063602 (2002).


\bibitem{Hadzibabic2016}
C. Eigen, A.L. Gaunt, A. Suleymanzade, N. Navon, Z. Hadzibabic, R.P. Smith, arXiv:1609.00352

\bibitem{Fattori2008}
M. Fattori, C. D'Errico, G. Roati, M. Zaccanti, M. Jona-Lasinio, M. Modugno, M. Inguscio, and G. Modugno,
Phys. Rev. Lett. {\bf 100}, 080405 (2008).

\bibitem{Shotan2014}
Z. Shotan, O. Machtey, S. Kokkelmans, and L. Khaykovich
Phys. Rev. Lett. {\bf 113}, 053202 (2014).


\bibitem{Franchini2015}
F. Franchini, A. Gromov, M. Kulkarni and A. Trombettoni,
J. Phys. A: Math. Theor. {\bf 48}, 28FT01 (2015).


\bibitem{Gamayun2015}
O. Gamayun, M. Semenyakin
J. Phys. A: Math. Theor. {\bf 49},  335201 (2016).

\bibitem{Polkovnikov2011}
A. Polkovnikov, K. Sengupta, A. Silva, and M. Vengalattore,
Rev. Mod. Phys. {\bf 83}, 863 (2011).

\bibitem{Cazalilla2011}
M.A. Cazalilla, R. Citro, T. Giamarchi, E. Orignac, M. Rigol,
Rev. Mod. Phys. {\bf 83}, 1405 (2011).

\bibitem{Goldstein2014}
G. Goldstein, N. Andrei,
Phys. Rev. A {\bf 90}, 043625 (2014).
%\bibitem{Streltsov2008}
%A.I. Streltsov, O.E. Alon, and L.S. Cederbaum,
%Phys. Rev. Lett. {\bf 100}, 130401 (2008).


\bibitem{Petrov2014}
D.S. Petrov, Phys. Rev. Lett. {\bf 112}, 103201 (2014).

\bibitem{Petrov2015}
D.S. Petrov, 
Phys. Rev. Lett. {\bf 115}, 155302 (2015).

\bibitem{Petrov2016}
D.S. Petrov, G.E. Astrakharchik,
arXiv:1605.07585


\bibitem{Kadau2016}
H. Kadau, M. Schmitt, M. Wenzel, C. Wink, T. Maier, I. Ferrier-Barbut, T. Pfau,
Nature {\bf 530}, 194 (2016).

\bibitem{Ferrier2016}
I. Ferrier-Barbut, H. Kadau, M. Schmitt, M. Wenzel, and T. Pfau,
Phys. Rev. Lett. {\bf 116}, 215301 (2016).



\bibitem{Chomaz2016}
L. Chomaz, S. Baier, D. Petter, M. J. Mark, F. W\"achtler, L. Santos, F. Ferlaino, arXiv:1607.06613

\end{thebibliography}

\end{document}